\documentclass{aa}  
\usepackage{graphicx}
\usepackage{txfonts}
\def\({\left(}
\def\){\right)} 
\def\[{\left[}
\def\]{\right]}

\begin{document}
\title{An inner disk below the ADAF: \\the intermediate spectral state of 
black hole accretion}

%   \subtitle{}

\author{B.F. Liu \inst{1}, F. Meyer \inst{2}, and E. 
Meyer-Hofmeister \inst{2} }
\offprints{Emmi Meyer-Hofmeister}
\institute
{National Astronomical Observatories/Yunnan Observatory, Chinese
Academy of Sciences, P.O. Box
110, Kunming 650011, China\\
\email{bfliu@ynao.ac.cn}
\and
Max-Planck-Institut f\"ur Astrophysik, Karl-
Schwarzschildstr.~1, D-85740 Garching, Germany\\
\email{frm@mpa-garching.mpg.de}
; emm@mpa-garching.mpg.de}

\date{Received: 13 April 2006 / Accepted: 1 June 2006}

% \abstract{}{}{}{}{} 
% 5 {} token are mandatory

\abstract
{{\it Aims:} The hard and soft spectral states of black hole accretion are
  understood as connected with ADAF accretion (truncated disk) and
  standard disk accretion, respectively. However, observations indicate the
  existence of cool gas in the inner region at times when the disk 
  is already truncated outside. We try to shed light on these not
  yet understood intermediate states.

   {\it Methods:} The disk-corona model allows to understand the spectral state
  transitions as caused by changes of the mass flow rate in the disk
  and provides a picture for the accretion geometry when disk
  truncation starts at the time of the soft/hard transition,  
  the formation of a gap in the disk filled by an advection-dominated
  flow (ADAF) at the distance where the evaporation is maximal. We
  study the interaction of such an ADAF with an inner thin disk below.

   {\it Results:} We show that, when the accretion rate is not far below the
   transition rate, an inner disk could exist below an ADAF, leading to 
  an intermediate state of black hole accretion.
   
}
\titlerunning {An intermediate state of black hole accretion}
\maketitle
%
%________________________________________________________________

\section{Introduction}
It is now widely accepted that the hard and soft spectral states of
black holes correspond to accretion in form of an advection-dominated 
flow (ADAF) and a thin disk, respectively (Esin et al. 1997). 
But observations also show an intermediate state which often appears in 
transitions between these two states.
   
The hard states are understood as a consequence of interaction of the
corona and the underlying disk which yields evaporation of gas from
the disk to the hot coronal flow, and leads to a
truncation of the accretion disk at some distance (Meyer et
al. 2000b). Inside this 
truncation radius there is only a hot, optically thin flow. The soft state, 
on the other hand, is explained as the optically thick, standard accretion 
disk extending down to the last stable orbit, revealing itself in the 
characteristic multi-temperature blackbody spectrum.

But observations also show an intermediate state which appears often in 
transitions between these two states (McClintock and Remillard 
2006). After the soft/hard transition 
the spectra are not always solely produced by an optically thin
accretion flow, the occurrence of reflection and a Fe K$\alpha$ line 
indicate cool matter in the inner region (\.Zycki et al. 1998). 
These intermediate states often persist for times significantly longer
than the viscous timescale of the disk and can thus not be 
explained only by the fading or build-up of a disk during transitions
between the two states.
We here address the accretion geometry in the intermediate
state. (Physically different is the so-called very high/intermediate
spectral state (discussion by Fender et al. 2004, model of disk
fragmentation Meyer 2004.) 

After a short description of the disk corona model in
Sect. 2 we analyze in Sect.3 the physics of an ADAF above a disk. 
One of the key 
questions is whether heat can be drained from the upper ADAF and
radiated away. We investigate the processes in the two-temperature
regime which extends over most of the vertical height, as well as those 
in the thin layer above the disk surface where ion and electron
temperature couple. In Sect. 4 we derive under which conditions
condensation of
matter from the ADAF into the disk occurs, the necessary
process to allow an inner disk to survive. In Sect.5 we 
discuss how this picture can be related to the observed 
intermediate states.

\section{Accretion geometry - Spectral states}
The key feature to understand the accretion geometry in the hard state
is the evaporation of gas from the disk, feeding the hot flow. 
The disk corona model (Meyer et al. 2000a) describes the  
interaction between the hot corona and the cool disk below
via energy and mass exchange.
Numerical calculations (Meyer et al. 2000a; Liu et al. 2002) show that
the evaporation rate reaches a maximum of about 1 percent of the
Eddington accretion rate at a distance of several hundred
Schwarzschild radii. Such a character, the maximum, provides a 
physical explanation for the occurrence of hard and soft states in
X-ray binaries and the transitions between them (Meyer et al. 2000b): If the 
mass flow in the disk is below this maximum value, as usual in 
quiescence, the disk is truncated at the distance where all matter is 
evaporated, leaving inside a pure coronal flow/ADAF, which produces the hard
spectrum. If the mass flow increases as during rise to an
outburst the edge of the disk moves inward. 
If the mass flow in the disk becomes higher than the maximal
evaporation rate, the disk can extend down to the last stable orbit and the
spectrum is soft. During outburst decline the opposite happens,
the inner edge of the disk retreats, in agreement
with the observations.  

\begin{figure}\label{schematic}
   \centering
  \includegraphics[width=8.cm, viewport=0 35 552 510, clip=]{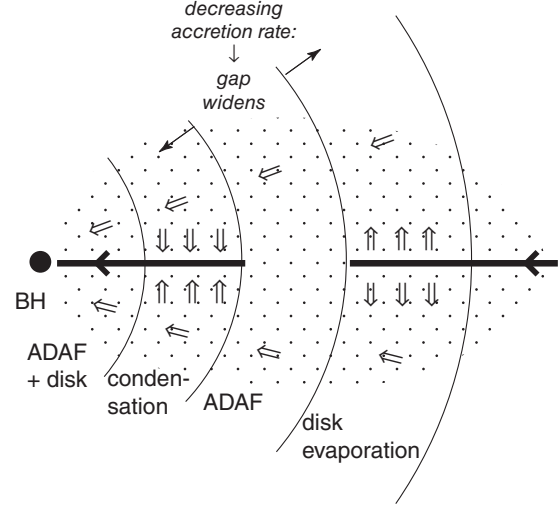}
   \caption{Schematic drawing of the accretion flow in disk and
   corona, a gap formed due to evaporation}   \label{f:vert}
    \end{figure}

In spite of this clear distinction of hard and soft states in the
standard picture a decrease of the mass flow
rate can
cause a more complicated accretion geometry, as a gap in the disk at
that distance where the evaporation efficiency is maximal
(Fig. \ref{f:vert}). Then the inner disk is affected by 
accretion towards
the center and evaporation of matter diffusing outwards. 
The question is: Can despite these processes a weak
interior disk exist during the hard, ADAF-dominated state? 

\section{Physics of an ADAF above a disk}
\subsection{Consequences of thermal conduction}
In a pure ADAF, almost all the accretion energy is stored in ions. Therefore, 
the ion temperature is near virial, the flow has a high vertical
extension, the density is low (review Narayan et al. 1998). The
coupling between electrons and ions is very poor and radiation is
quite weak. The situation is different if there is a disk below the
ADAF. Due to the large temperature difference between hot gas and
cool disk, thermal conduction results in cooling of
electrons. While the ion temperature is not much affected the electron
temperature drops with height until at some $T_e=T_{\rm cpl}$ coupling between
ions and electrons becomes efficient and from then on ion and electron
temperatures are the same, $T_i=T_e$. This 
constitutes a huge drop in ion temperature and causes a large
increase in density as the pressure in the 
vertical extent keeps almost constant. As a consequence, Bremsstrahlung 
becomes much more important than in a typical ADAF. 
If the pressure exerted by the ADAF is large enough, 
Bremsstrahlung can be so efficient that all the heat 
drained from the ADAF by thermal conduction is radiated away in this
layer. As in the standard disk corona model,
this decides on evaporation and condensation of gas from and to the disk. 

\subsection{Properties of the hot flow above a disk}
For our analysis we use the
solutions of self-similar advection-dominated flows by
Narayan \& Yi (1995). The properties of these flows scale with 
black hole mass, mass flow rate and distance from
the black hole. They further depend on the viscosity and
the assumed magnetic field strength. Following Narayan et al. (1998)
the magnetic pressure is written as   
\begin{equation}\label{p_mag}
p_{\rm m}=(1-\beta)\rho{c_{\rm s}}^2
\end{equation}
with $\beta$ ratio of gas pressure
to total pressure, $\rho$ density and $c_{\rm s}$
isothermal sound speed. Recent shearing box simulations of
turbulence driven by the magnetorotational instability in a
collisionless plasma by Sharma et al. (2006) yield $\beta$ values
around 0.8. We take this value for our analysis. As ratio of specific heats of
the magnetized plasma we take $\gamma=(8-3\beta)/(6-3\beta)$ (Esin
1997) though the true value would require a more detailed analysis.

For the chemical abundance a hydrogen mass fraction of 0.75 was used.
The solutions for pressure, electron number density, viscous dissipation of
energy per unit volume $q^+$ and isothermal sound speed are 
\begin{eqnarray}\label{scaled}
p&=&1.87\times 10^{16}\alpha^{-1}m^{-1}\dot m r^{-5/2} 
\rm{g cm^{-1} s^{-2}} \nonumber ,\\
n_e& =&5.91\times10^{19}\alpha^{-1}m^{-1}\dot m r^{-3/2} \rm{cm^{-3}} \\
q^+&=& 2.24\times10^{20}m^{-2}\dot m r^{-4}\rm{ergs cm^{-3} s^{-1}}
\nonumber,\\
c_{\rm s}^2&=&1.67\times 10^{20} r^{-1} \rm{cm^2 \,s^{-2}} \nonumber.
\end{eqnarray} 
where $\alpha$ is the viscous coefficient,
$m$ the black hole mass in units of
solar mass $M_ \odot$, $\dot m$ the mass flow rate in units of 
Eddington accretion rate $\dot M_{\rm Edd}=1.39\times 10^{18}m\,{\rm
g/s}$, and $r$ the radius in units of Schwarzschild radius 
$R_{\rm S}=2.95\times 10^5 m
\,\rm{cm}$.
The ion number density is $n_i=n_e/1.077$.
In an ADAF ion and electron temperatures $T_i$ and $T_e$
closely follow  
\begin{equation}
T_i+1.077 T_e =1.98\times10^{12} r^{-1}\rm K 
\end{equation}
and, if $T_e$ is much smaller than $T_i$ this value can be taken for
$T_i$ alone. 
Besides viscous heating $q^+$ compressive heating
$q^c=\frac{1}{(1-\beta)}q^+$ (Esin 1997) is important. 
>From the scaled values we determine the vertical 
conductive heat flux $F_c$ and the temperature at which ions and
electrons couple.

The electron temperature has practically no influence on the dynamics
of the flow and the thermodynamics of the ion gas and is
subrelativistic, $kT_e <m_ec^2$ ($k$ Boltzmann constant, $m_e$
electron mass, $c$ speed of light). The energy
transfer from ions to electrons is given by Stepney (1983). Since in
the two-temperature advection-dominated hot flow the ions generally are at a
much higher temperature than the electrons a simplified 
formula (Liu et al. 2002) can be used 
\begin{eqnarray}\label {qie}
q_{ie} & = & 3.59\times 10^{-32} {\rm{g cm^{5} s^{-3} deg^{-1}}}
\, n_e n_i T_i 
{\left(\frac{k T_e}{m_e c^2}\right)}^{-\frac{3}{2}} 
  \nonumber \\
       & = & 1.05\times 10^{35} {\rm{ g cm^{-1}s^{-3}deg^{3/2}}} 
T_e^{-3/2} \alpha^{-2}m^{-2} \dot m^{2}r^{-4}. 
\end{eqnarray}

Cooling of electrons occurs through a
variety of channels. Bremsstrahlung and synchrotron (-Compton) radiation 
(depending on the strength of the magnetic field) limit the electron 
temperature in the inner ADAF region. The new element 
is that an underlying disk drains heat. This is a new cooling process
that competitively limits the peak electron temperature.
This process determines the drop of electron temperature with 
height $z$ above the disk. The downward heat
flux increases from $F_{\rm c}$=0
at large height to a saturation value at the bottom where ion and
electron temperatures finally couple, at $z=z_{\rm{cpl}}$. For our approximation
we take $z_{\rm{cpl}}$ as small compared to the extent of the ADAF
(in our example of order 10\%).
The heat flux follows from the two relations
\begin{eqnarray}\label{heatflux}
F_{\rm c}&=&-\kappa_0 T_e^{5/2}dT_e/dz,\\
\frac{dF_{\rm c}}{dz}&=&-q_{ie}(T_e)
\end{eqnarray}
by integration over this temperature interval. We get
\begin{equation}\label{Fc2}
{F_{\rm c}}^2=\kappa_0 (K n_i n_e T_i)  
{({T_{\rm m}}^2-{T_e}^2})
\end{equation}
with $K= 1.64\times 10^{-17} \rm {g cm^{5}s^{-3}deg^{1/2}}$.
The integration of $z$ over $T_e$ (Eq.{\ref{heatflux}}) 
gives the relation between the maximal electron temperature $T_{\rm m}$ and $z_{\rm m}$, the height at which  $T_{\rm m}$ is reached. This yields 
\begin{equation}\label{T_0}
{T_m}^{5/2} = 1.39\, z_m \,{(K n_i n_e T_i/ \kappa_0)}^{1/2}.
\end{equation}
For the height $z_{\rm {m}}$ we take the vertical 
scaleheight for $n_e n_i$ which is $1/\sqrt{2}$ of the density
scaleheight $c_{\rm s}/\Omega_{\rm K}$ ($\Omega_{\rm K}$ Kepler angular
velocity) of  Narayan et al. (1998).
The value of $T_{\rm{m}}$ finally yields the heat flux near the bottom 
($z=z_{\rm{cpl}}$) as
\begin{equation}\label{F_c}
F_{\rm c}^{\rm{ADAF}}=-
\kappa_0^{3/10}\, {(K n_i n_e T_i)}^{7/10} 
{(1.39\, z_{\rm{m}})}^{2/5}
\end{equation}

\subsection{Coupling of ion and electron temperature}
As pointed out in Sect. 3.1 only if the electron temperature has
dropped sufficiently ions and electrons start to couple. This
requires that the heat 
transfer from ions to electrons balances viscous and
compressive heating, $q_{ie}=q^+ + q^c$. The coupling
temperature follows from Eq.({\ref{scaled}}) and Eq.({\ref{qie}}). 
Since the coupling occurs at very low height 
the density there is higher than the height averaged
value (Eq. {\ref{scaled}), by a factor is $2/{\sqrt \pi}$ (vertical
density distribution $n=n_0\,\rm{exp}(-{z^2}/{H^2})$. This yields the 
coupling temperature as
\begin{equation}\label{T_cpl}
T_{\rm {cpl}}=1.98\times 10^9 \alpha^{-4/3}\dot m^{2/3}= 1.24\times 10^9 K,
\end{equation}
the numerical value for $\alpha$=0.2 and $\dot m$=0.02 (see Sect.4). 
The sudden decrease of ion temperature leads to a sudden
increase of ion and electron density, and the electrons 
radiate efficiently.

\subsection{The radiating layer close to the disk surface}
The parameters of the ADAF determine what happens in the layer between 
coupling and disk surface.
Different regimes are possible, depending on 
the pressure in the ADAF.  Case (1): If the pressure is high, the 
density in the lower layer also is high. Then the
conductive flux drained from the ADAF is already efficiently radiated
away at some height before the disk surface is reached. Radiative
cooling must be additionally balanced by heat released from gas
condensing  
into the disk. Case (2): If the pressure is low, only part of the
conductive flux can be radiated away, the remaining part will heat up 
cool disk matter, i.e. matter evaporates from the disk to the corona. 
A limiting case in between, case (3), occurs if the ADAF pressure 
allows the thermal heat flux drained from the ADAF to be 
radiated away exactly at the disk surface.
Then there is neither condensation nor evaporation, no mass exchange 
between the disk and the ADAF. 

\section{Condensation of gas from the ADAF into the inner disk?}
>From the energy balance in the radiating layer we derive the condition 
for the different
cases. We use here a simplified form of the energy equation,
keeping only the dominant contribution of internal heat, pressure
work, and thermal conduction, together with Bremsstrahlung cooling 
$n_e n_i \Lambda(T)$

\begin{equation}\label{energy}
\frac{d}{dz} \left[\dot m_z \frac{\gamma}{\gamma-1} \frac{\Re T}{\mu} + 
F_c \right] = -n_e n_i \Lambda(T).
\end{equation}
with $\Re$ gas constant, $\mu$ molecular weight, $\dot m_z$
vertical mass flow rate between ADAF and disk per unit area.  
We express density by temperature and (constant) gas pressure
$\beta\,p$. This value is taken at the bottom of the ADAF
region, and, as for the density, is slightly higher than the 
vertical mean pressure (Eq.(\ref{scaled})). Assuming free-free 
radiation for $T_e \ge 10^{7.5}\rm{K}$,  $n_e n_i \Lambda(T)$ becomes 
$\frac{0.25}{k^2}{(\beta\,p)}^2 bT^{-3/2}$ with
$b=10^{-26.56} \rm{g\,cm^{5} s^{-3} deg^{-1/2}}$ (Sutherland \& Dopita 1993). 
For formal simplicity we use this law also for smaller $T$. The
justification is that because of the very steep temperature profile
below $10^{7.5}$K such regions contribute only a negligible amount to
cooling of this layer (Liu et al. 1995).
Contributions of gravitational energy release, frictional heating, and 
side-wise advection of mass and energy can be neglected for small
extent of this layer. Likewise kinetic energy is negligible since at
the high density the flow is highly subsonic.

To solve the second-order differential equation Eq.(\ref{energy}),
we change to $T$ as independent variable and define a new variable 
$g(T)\equiv \kappa_0T^{3/2}dT/dz=-F_{c}/T$. We now get the
first-order differential equation
\begin{eqnarray}\label{gT}
g \frac{dg}{d\ln T} &=& \frac{0.25\beta^2 p^2}{k^2} \kappa_0 b
 +  \dot m_z \frac{\gamma}{\gamma-1} 
\frac{\Re}{\mu}g - g^2 
\nonumber \\
&=& - (g-g_1)(g-g_2),
\end{eqnarray}
with 
\begin{equation}\label{g_1}
g_1={\dot m_z\over 2} {\gamma\over \gamma-1} {\Re\over \mu}+
\sqrt{\left({\dot m_z\over 2} {\gamma\over \gamma-1} {\Re\over
\mu}\right)^2
+\frac{0.25\beta^2 p^2}{k^2} \kappa_0 b}
\end{equation}
and $g_2$ differing from $g_1$ only by the sign of the square root.

This equation has to be solved with the upper boundary
at $T=T_{\rm{cpl}}$, where ${F_c}^{\rm{ADAF}}$ is the heat flux
coming down from to the two-temperature ADAF region, and the
lower boundary at $T=0$, the disk surface, with $F_c=0$. 
$\dot m_z$ is the Eigenvalue. The only solution that fulfills both
boundary conditions is the singular solution $g(T)=g_1$. We get 
\begin{eqnarray}\label{C}
\dot m_z & = & \frac{\gamma-1}{\gamma} 
\frac{- {F_c}^{\rm {ADAF}}} {\Re T_{\rm {cpl}} 
/{\mu}}
\[1- C \], \\
C & = & \kappa{_0} b 
\(\frac{0.25\beta^2 p^2}{k^2}\)
\(\frac{T_{\rm {cpl}}}{{F_c}^{\rm {ADAF}}}\)^2 \nonumber
\end{eqnarray}
The sign of the square bracket decides on evaporation or condensation: If
$C$ is less than 1, the radiation losses are
too weak, $\dot m_z$ is positive and the heat flow is used up to heat
the evaporating gas to the boundary temperature $T_{\rm cpl}$.
If, on the other hand, $C>1$,
then the gas in this layer is efficiently cooled, sinks down and
condenses. Note that at the coupling boundary for a rising flow 
additional heat is used up to heat the ions to the ADAF value $T_i$. 
This diminishes ${F_c}^{\rm{ADAF}}$ in Eq.(\ref{C}). The opposite
holds for a condensing flow. This limits the use of Eq.(\ref{C}) to
the neighborhood of $C=1$, the borderline case.

The value of $C$ scales with $\alpha$, $\dot m$ and $r$, but not with
$m$. This shows that the same situation holds for supermassive and
stellar black hole accretion. It
depends on the  gas pressure fraction $\beta$. For $\beta$=0.8 we get 
\begin{equation}\label{ccrit}
C=0.96\, \alpha^{-28/15} \dot m ^{8/15} r^{-1/5}
\end{equation}
We take $\alpha$=0.2 (corresponding to $\alpha$=0.3 in
Shakura-Sunyaev notation). The intermediate states appear in
connection with the spectral state transitions. Maccarone (2003) found
from a detailed investigation of these transitions in X-ray
binaries $\dot m=0.02$.
With these values $C$ becomes 1 at the distance of about 80 
Schwarzschild radii. Note significant dependence on the
value of $\beta$. The important outcome is the dependence of the value 
$C$ on $\dot m$ and $r$, which 
allows to derive the conditions for the existence of \bf{cool
matter} \rm {below the ADAF.

\section{Understanding the intermediate state}
The dependence of $C$ on mass flow rate and distance yields the
following picture. If during outburst decline the mass flow rate in the disk
decreases to a value equal to the maximal evaporation rate the disk
breaks up where the evaporation is most efficient. A gap
forms, inside a cool disk still exists, an ADAF above it. Only if 
condensation is possible this disk can survive.
Our analysis predicts:
If the mass flow rate declines slowly, from the
transition rate to the minimal rate for condensation to
occur ($\dot m\approx 0.006$, $C=1$ at the last
stable orbit), an intermediate spectral state appears, caused
by a sustained inner disk underneath an ADAF. Note that due to the low
recondensation rate a soft contribution to the spectrum is weak. 
If the rate drops quickly,
almost no condensation occurs, the disk disappears as a consequence of 
accretion and evaporation without new mass supply from the ADAF.

Kalemci et al. (2004) studied intermediate states of several
X-ray transients. Our analysis suggests that these states are related 
to a phase where the disk breaks up, but inner cool matter remains by
recondensation from an ADAF that has formed. 
But comparison with observations still requires further analysis,
in particular of processes which tend to limit the inward extent of the 
remaining cool matter.

During rise to outburst the situation is different, the inner
edge of the disk moves inward within a viscous time, inside only an
ADAF. 

If the mass flow rate in the disk is always well below the
maximal evaporation rate the disk is truncated all the time,  
no condensation occurs, since no cool matter exists inside the
truncation, no thermal conduction can appear.
This picture is supported by observations of XTE J1118+480, a system
which stays in the hard state even during outburst (Esin et
al. 2001).

\section{Conclusions}
We introduce a new feature of accretion flows onto black holes, the
advection-dominated accretion flow affected by thermal conduction to a
cool disk underneath. Our analysis shows that in the inner region
condensation of matter from the hot flow into the disk is
possible. 
This allows the existence of cool matter together with an already
formed ADAF as indicated by the observed reflection and Fe K$\alpha$
line in the intermediate spectral state.

\begin{acknowledgements}
The authors acknowledge support by
the National Natural Science Foundation of China (NSF-10533050)
and the BaiRenJiHua program of the Chinese Academy of Sciences.
\end{acknowledgements}
}

\end{document}